\documentclass[american,preprint,pra,aps,amssymb,amsfonts,amsmath]{revtex4}
\usepackage{amssymb,amsmath}
\usepackage{tabularx}
\usepackage{dcolumn}
\usepackage{multirow}
\usepackage{array}
\usepackage{graphicx}
\newcommand{\mc}[3]{\multicolumn{#1}{#2}{#3}}

\begin{document}
\title{On the Cholesky Decomposition for electron propagator methods:\\
General aspects and application on C$_{60}$}
\date{\today}
\author{Victor P. Vysotskiy}
\email{Victor.Vysotskiy@pci.uni-heidelberg.de}
\author{Lorenz S. Cederbaum}
\email{Lorenz.Cederbaum@pci.uni-heidelberg.de}
\affiliation{Theoretical Chemistry, Institute of Physical Chemistry at
Heidelberg University, Im Neuenheimer Feld 229, 69120 Heidelberg, Germany}

\begin{abstract}
To treat the electronic structure of large molecules by electron propagator methods we developed a parallel computer program called $\mathrm{P{-}RICD}\Sigma$. The program exploits the sparsity of the two-electron integral matrix by using Cholesky decomposition techniques. The advantage of these techniques is that the error introduced is controlled only by one parameter which can be chosen as small as needed. We verify the tolerance of electron propagator methods to the Cholesky decomposition threshold and demonstrate the power of the $\mathrm{P{-}RICD}\Sigma$ program for a representative example (C$_{60}$). All decomposition schemes addressed in the literature are investigated. Even with moderate thresholds the maximal error encountered in the calculated electron affinities and ionization potentials amount to a few $m$eV only, and the error becomes negligible for small thresholds.
\end{abstract}

\maketitle

\section{I. Introduction}

In all advanced quantum chemistry calculations the evaluation of the huge number of two-electron repulsion integrals (ERIs) consistutes the major computational obstacle. It is thus not surpising that one searches for efficient methods to accurately approximate these integrals. Most known is the method of resolution of identity (RI), see, e.g. the reviews~\cite{Kendall,ADC2} and references therein, which has been widely used to compute energies and other properties of molecules. A few decades ago Beebe and Linderberg have published an original paper about reducing computational efforts of ab initio methods by using Cholesky decomposition (CD) of the two-electron integral matrix~\cite{BeebeLind}. Within the CD method the accuracy of representation of the exact ERIs is controlled only by a single parameter, the so-called Cholesky decomposition threshold $\delta$. By construction, it provides an upper bound to the absolute difference between an exact ERI and its approximated value. One of the key advantages of the CD approach is that the error introduced to the ERIs can be made as small as needed. Recently, the CD approach has been successfully combined with MP2 perturbation theory~\cite{ALG1,PT2,PT3}, with CASSCF configuration interaction methods~\cite{CAS1,CAS2,CAS3} and with the CC2 linear response theory~\cite{CC2LR1,CC2LR2}. By using the CD approach one can speed up electron correlated calculations up to a few hundred times, thus enabling applications to large quantum systems (many atoms and many basis functions). Moreover, within the CD method one can perform correlated calculations of systems, which can not be attacked using ordinary conventional techniques.

Our goal is to study the performance of CD in electron propagator calculations. We would like to mention that Flores-Moreno and Ortiz have investigated  and applied the RI method in the context of electron propagator theory~\cite{RIEP}. Electron propagator theory (EPT) has been proven to be a powerful tool for investigating the electronic structure of bound and unbound, metastable, states. Indeed, the EPT is widely used for the direct calculation of vertical ionization and attachment energies~\cite{CEDER77,CEDER84,DEL1,DEL2,OrtizR,OrtizB,CEDER98,OHRN}, and more recently also of widths and positions of short-lived electronic states (resonances)~\cite{SANTRA,CAP1}. In particular, methods have been formulated as a combination of the EPT and a stabilization technique and successfully used for describing resonant states of anions~\cite{CAP2,CPES1,CAP3}.

Normally, the error of computed total energies (using standard methods like e.g. SCF, MP2, CASSCF, CASPT2) or response properties
introduced by the CD approximation has the same order of magnitude as $\delta$ ~\cite{ALG1,ALG2,LKCD,CC2LR2}. In contrast, Beebe and Linderberg made the assumption that electron propagators should be tolerant to the Cholesky decomposition threshold: "$\ldots$ for propagator methods which directly compute energy differences, for an accuracy of $10^{-5}$ in the transformed two-electron integrals will be more than adequate for most purposes
$\ldots$"~\cite{BeebeLind}. In the last few years substantial progress has been made in further optimization of the CD method. In particular,
a new, so-called \textit{atomic} CD version has been developed by F. Aqualiante et al.~\cite{aCD1}. The main goal of our present work is to investigate the feasibility of using the CD approaches including the atomic version for electron-propagator methods.

\section{II. THEORY}
\subsection{Factorization of two-electron repulsion integrals}
Let $\lbrace \chi_{\mu} \left(r\right);1{\le}\mu{\le}{N}\rbrace$ denote a set of real spatial  basis functions and $\mathfrak{R}{=}\lbrace\rho_{I}\left(r\right)=\chi_{\mu}\left(r\right)\chi_{\nu}\left(r\right);I\equiv(\mu,\nu),1{\le}\mu{\le}\nu{\le}N,I{\le}{N{\left(N+1\right)}}/2\rbrace$ denote the corresponding one-electron product density set.
One of the basic quantities in quantum chemistry is a two-electron repulsion integral (ERI):
\begin{multline}
\label{integral}
\left(\mu \nu \lvert \lambda \sigma \right)=\iint\chi_{\mu}\left(r_1\right) \chi_{\nu}\left(r_1\right) \frac{1}{r_{12}} \chi_{\lambda}\left(r_2\right)
\chi_{\sigma}\left(r_2\right)d\boldsymbol{\mathrm{r}}_1d\boldsymbol{\mathrm{r}}_2= \\ \iint \rho_{I}\left(r_1\right)  \frac{1}{r_{12}}\rho_{J}\left(r_2\right)
d\boldsymbol{\mathrm{r}}_1d\boldsymbol{\mathrm{r}}_2=\left(\rho_I \lvert V \lvert \rho_J \right)
\end{multline}
where
$r_{12}=\lvert\boldsymbol{\mathrm{r}}_1-\boldsymbol{\mathrm{r}}_2\lvert$.

Since  ERI is a four-indexed quantity, both the number of distinct ERIs and the time required for their evaluation increase as $N^4/8$. $O\left(N^4\right)$ scaling forms one of the basic obstacles to ab initio calculations with a finite basis set. The Density Fitting (DF)~\cite{DF1,DF2,DF3,DF4,DF5_GS} or Resolution of Identity (RI) ~\cite{RI,TJLee,Kendall}  and Cholesky Decomposition (CD) ~\cite{BeebeLind} methods are very efficient approaches to reduce scaling. The key idea in all of these methods is to factorize
an ERI into a product form of three-indexed quantities ~\cite{Pseudospectral}:
\begin{equation}
\label{factorization}
\left(\mu \nu \lvert \lambda \sigma\right)=\sum_{K=1}^{M}A_{\mu\nu}^KA_{\lambda\sigma}^K\thinspace.
\end{equation}
The scaling now is $O\left(N^3\right)$. This factorization also significantly reduces I/O, memory demands and also reduces the scaling of the atomic orbital (AO) to molecular orbital (MO) transformation from $O(N^5)$ to $O(N^4)$ ~\cite{BeebeLind}. The last issue is crucial for post-HF methods. Denote  the MO expansion coefficients matrix by \textbf{C}, then
an ERI in MO representation can be calculated by exploiting the same factorization~(\ref{factorization}):
\begin{equation}
\label{aomo}
\left(pq\vert rs\right)=\sum_{K=1}^{M}B_{pq}^KB_{rs}^K\thinspace,
\end{equation}
where $p,q,r,s$ denote MO indices and $B_{pq}^K$:
\begin{equation}
\label{transformation}
B_{pq}^K=\sum_{\mu,\nu} C_{\mu p} A_{\mu\nu}^KC_{\nu q}\thinspace.
\end{equation}

DF/RI and CD approaches vary in the generation of the intermediate $A_{\mu\nu}^K$. In order to demonstrate the interplay between DF/RI and CD techniques we would like to derive the given factorization ~(\ref{factorization}) by L\"{o}wdin's inner projection technique~\cite{Lowdin,LowdinLA}\quad(see also ~\cite{BeebeLind}). Let us introduce the expansion basis set of functions $\widetilde{\mathfrak{R}}{=}\lbrace{ \widetilde{\rho_I}\left(r\right);1{\le}I{\le}M}\rbrace$ in the domain of the Coulomb operator  $\mathbf{V}{=}r_{12}^{-1}$ and construct a projection operator \textbf{P} as follow:
\begin{equation}
\label{projector_general}
\mathbf{P}=\sum_{K,L} \mathbf{V}^{1/2} \lvert \widetilde{\rho_K}) V^{-1}_{KL} (\widetilde{\rho_L} \lvert \mathbf{V}^{1/2}\thinspace,
\end{equation}
where 
$V^{-1}_{KL}$ is the $KL$th element of the inverse $\mathbf{V}^{-1}$ and  $V_{KL}=\left(\widetilde{\rho_K} \lvert V \lvert \widetilde{\rho_L}\right)$. The inner projection of the Coulomb operator \textbf{V} on \textbf{P} takes on the form
\begin{equation}
\label{inner_general}
\mathbf{\widetilde{V}}=\mathbf{V}^{1/2}\mathbf{P}\mathbf{V}^{1/2}=\sum_{K,L} \mathbf{V}\lvert \widetilde{\rho_K}) V^{-1}_{KL} ( \widetilde{\rho_L} \lvert \mathbf{V}\thinspace,
\end{equation}
and provides the following approximation to an ERI:
\begin{equation}
\label{int_approx}
\left( \rho_{I} \lvert V\lvert \rho_{J}\right)=( \rho_{I} \lvert \widetilde{V} \lvert \rho_{J})+D_{IJ} \approx \sum_{K,L} \left( \rho_{I} \lvert V \lvert \widetilde{\rho_K} \right) V^{-1}_{KL} \left( \widetilde{\rho_L} \lvert V \lvert \rho_{J} \right)\thinspace,
\end{equation}
where
\begin{equation}
\label{error_approx}
0\le( \rho_{I} \lvert V-\widetilde{V}\lvert \rho_{J})=D_{IJ}\thinspace,
\end{equation}
is the error introduced by the incompleteness of the expansion basis set. The double sum in the right hand side of Eq.~(\ref{int_approx}) is the famous \textbf{V}-approximation which comes from the RI methodology ~\cite{RI}.
Noting that since $\mathbf{V}$ is a positive definite operator and $\mathbf{\widetilde{V}}$ is the lower bound to it, $D_{IJ}$ satisfies the Schwarz inequality:
\begin{equation}
D_{IJ}{\le}D_{II}^{1/2}D_{JJ}^{1/2}\thinspace.
\end{equation}

The difference between CD and DF/RI methods are the generation of the expansion basis set $\widetilde{\mathfrak{R}}$ or, in other words, they differ in the fitting of the original density product set $\mathfrak{R}$. Here, there are two possibilities to construct expansion basis sets: using a native subset of the original product density set or using an external auxiliary basis set. The CD method forms an expansion basis set from the original $\mathfrak{R}$ ($\widetilde{\mathfrak{R}}\subset\mathfrak{R}$). It exploits the linear dependence of the original  product density set in the Coulomb metric ~\cite{BeebeLind,LD2,LD3}. Thus, the CD method is the ab initio density fitting by construction (by design). For constructing a linear independent subset of $\mathfrak{R}$ it is convenient to apply the classical Gram-Schmidt (CGS) method  using $\mathbf{V}$ as the weighting factor:
\begin{equation}
\label{gs_coulomb}
\widetilde{\rho_K}=n_K^{-1/2}\biggl[{\rho_K}-\sum_{I=1}^{K-1}\widetilde{\rho_I}\left( \widetilde{\rho_I} \lvert V \lvert \rho_K\right) \biggr]\thinspace,
\end{equation}
where the normalization constant $n_K$ is given by
\begin{equation}
n_K=\left( \rho_K \lvert V \lvert \rho_K \right)-\sum_{I=1}^{K-1}{\left( \widetilde{\rho_I} \lvert V \lvert \rho_K\right)}^2\thinspace.
\end{equation}

This recursive procedure is continued as long as the norm of the new vector $\rho_{M_{\delta}}$ (at step $\rm{M}_{\delta}$) is greater than the threshold for linear dependency $\delta$, i.e. $n_K\ge\delta$, or equivalently,
$D_{KK}\le\delta$~\cite{SORT}. The threshold $\delta$ is known as the \textit{Cholesky decomposition threshold} and the resulting non-redundant subset is known as the \textit{Cholesky basis}. As a rule of thumb, if $\delta=10^{-p}$, then~\cite{PowerCD}:
\begin{equation}
 pN \le M_{\delta}\le\left(p+1\right)N\thinspace.
\end{equation}

It is interestingly to note that the same idea of using the CGS procedure for generating expansion basis set has been suggested by Whitten in the frame of the DF approach about 30 years ago ~\cite{DF5_GS}. By taking into account that $\left( \widetilde{\rho_I} \lvert V \lvert \widetilde{\rho_J} \right){=}{\delta}_{IJ}$ and returning to the original set of pair of indices, the final expression of Eq.~(\ref{int_approx})  in the Cholesky basis takes on the form: \\
\begin{equation}
\label{CD} 
\left( \mu\nu\lvert \lambda\sigma \right) \approx\sum_{K=1}^{M} \left( \mu \nu \lvert \widetilde{\rho_K} \right)  \left( \widetilde{\rho_K} \lvert \lambda\sigma \right)\equiv\sum_{K=1}^{M} L_{\mu\nu}^K L_{\lambda\sigma}^K\thinspace,\thinspace M=min\bigl(M_{\delta},\left(\mu\nu\right)\mapsto I,\left(\lambda\sigma\right)\mapsto J\bigr)\thinspace,
\end{equation}
where the intermediate $L_{\mu\nu}^K$ is the projection of the $\vert\mu\nu)$ product on the $K$th vector of $\widetilde{\mathfrak{R}}$. The vector $\mathrm{\mathbf{L}^K}{=}\lbrace L_{\mu\nu}^K;1{\le}\mu{\le}\nu{\le}N\rbrace$ is known as \textit{integral table}. The number of ERIs needed for the generation of $M_{\delta}$ integral tables is $M_{\delta}N(N+1)/2$ ($O(N^3)$). In actual practice a modified Cholesky decomposition algorithm~\cite{ALG1,ALG2,ALG3} is used for calculating the intermediates $L_{\mu\nu}^K$ instead of CGS  because the latter is numerically unstable~\cite{CGSUNST}.

The Cholesky basis has three important features. First, it is optimal in the sense that it provides a rigorous upper bound for the value $D_{IJ}$,
i.e.:
\begin{equation}
\label{stricterr}
D_{IJ}\le\delta
\end{equation}
An important point that if $\rho_I$ and $\rho_J$  have been involved in the CGS orthonormalization process, then the corresponding integral is represented exactly~\cite{ALG3}, i.e.,
\begin{equation}
D_{IJ}=\epsilon\thinspace,
\end{equation}
where $\epsilon$ is the machine epsilon. This means that $M_{\delta}\left(M_{\delta}+1\right)/2$ integrals corresponding to
the product densities of $\widetilde{\mathfrak{R}}$ will be generated exactly (within machine precision) by Eq.~(\ref{CD}) ~\cite{BeebeLind,ALG3}.
Second, the Cholesky basis is continuous in the sense that by decreasing the threshold $\delta$ the inner projection $\mathbf{\widetilde{V}}$ smoothly becomes better and better and formally exact when asymptotically $\delta{\rightarrow}0$, i.e.:
\begin{equation}
\lim_{\delta\to0}\mathbf{\widetilde{V}}{=}\mathbf{V} \mbox{\rm {\quad and \quad}} \lim_{\delta\to0}D_{IJ}=\epsilon\thinspace.
\end{equation}
In the finite precision arithmetic this limit is reached with a Cholesky threshold less or equal to $10^{-10}$ ($\delta\le10^{-10}$).
Thus, by using the CD approximation the error introduced to ERI can be made as small as desired. This features is a great advantage of the CD method. Third, the Cholesky basis set can be constructed on the fly for any arbitrary AO basis set, even for non-standard ones. Such a flexibility is important when calculating short-lived electronic states, where it is essential to extend an original AO basis set by adding external sets of basis functions for describing the continuum wavefunction~\cite{Kaufmann,Nestmann,Cubic}. 

The disadvantage of the full-CD is the total price in the number of operations (multiplication, addition, square roots and array indexing is $\propto N^4)$. By construction, the Cholesky basis generally contains  both one-center and two-center one-electron densities which lead to the calculation of expensive four-center ERIs and, more important, to a dependence of the Cholesky basis on the molecular geometry~\cite{DiscCB}. To overcome these shortcomings a so-called atomic CD (aCD) modifications has been developed recently. In the aCD version the Cholesky basis contains only one-center product densities for each unique atom/AO basis set pair which reduces the computational effort considerably and leads to smooth potential energy surfaces~\cite{aCD1,DiscCB}. The aCD (\textit{atomic Cholesky}) basis set provides strict error control ~(\ref{stricterr}) on the one-center and two-center "Coulomb" ERIs but three- and four-centers integrals and "Exchange" two-center integrals may be affected by large errors~\cite{acCD1}. Unlike the full-CD approach, the accuracy of the aCD method  cannot be improved beyond a certain limit, which is weakly dependent on the decomposition threshold. Nonetheless, recently it has been shown that the aCD approach does not downgrade the accuracy to any significant degree and the introduced error is slightly larger than that of the original
full-CD~\cite{CAS1,aCD1,DiscCB}.

There is a further optimization of the aCD basis, the so-called atomic compact CD (acCD) basis set. The acCD is the result of removing the linear dependence among the primitive Gaussians ~\cite{acCD1}. As has been demonstrated recently in a serie of papers, the accuracy of aCD and acCD are very close to each other ~\cite{aCDCal,acCD1}. It is important to stress that the Cholesky basis sets are generated from first principles and are not biased to any quantum chemical method. For a detailed discussion on atomic CD we refer to the review ~\cite{aCDREW}. 

\uchyph=0

In contrast to the above mentioned \textit{ab initio} Cholesky basis sets, the DF/RI  auxiliary basis sets have been optimized to reproduce accurately certain specific quantities. The key point in the development of auxiliary basis sets is to provide a balance between the accuracy of the computed quantities and the numerical effort required. The balanced auxiliary basis set should fulfil two requirements. First of all its size should be only a few times (3-4) larger than the size of the original AO basis set. Second, the error introduced due to the DF/RI approximation should be at least one order of magnitude smaller than the error resulting from one-electron basis set incompleteness. Technically, an auxiliary basis set is optimized by minimizing the deviation between the exact target quantity and that calculated via the DF/RI approximation in a set of atomic and molecular calculations for a certain combination of pair AO basis set and level of theory. One of the best approved family of auxiliary basis sets is RI-X (X=J,JK,C), where X refers to the theoretical method used in the parameterization~\cite{RIJ1,RIJ2,RIJK1,RIJK2,RIC1,RIC2,RIC3}. 

In particular, the RI-JK family of auxiliary basis sets was developed to reproduce the HF coulomb and exchange energies:
\begin{equation}
 \label{SCF}
  \displaystyle E_{SCF}=\sum_{i,j}\left(\thinspace2(ii\lvert jj)-(ij\lvert ij\thinspace)\right)\thinspace,
\end{equation}
whereas RI-C was invented to reproduce the MP2 correlation energy:

\begin{equation}
 \label{MP2}
  \displaystyle MP2_C=\sum\limits_{\substack{i,j\\a,b}}\frac{\left(\thinspace 2(ai\lvert bj)-(aj\lvert bi)\right)\cdot(ai\lvert bj)}
{\varepsilon_i+\varepsilon_j-\varepsilon_a-\varepsilon_b}\thinspace,
\end{equation}
where $a,b$ denote virtual, and $i,j$ denote occupied HF orbitals. Later on H\"{a}ttig et al. demonstrated that the RI-C auxiliary basis sets also suitable for calculating
ground state and excitation energies at the CC2 level of theory~\cite{CC2,ADC2}. The typical errors produced by RI-X\- auxiliary basis sets for a target quantity are about a fraction of few tenths or even few hundredth of $m\mathrm{E_h}$ per atom. However, in the case of an inappropriate combination of AO and RI-X basis sets and level of theory the resulting error can increase drastically and be a few orders of magnitude larger than suggested in the original
papers~\cite{RIvsCD}. The other issue is the application of the DF/RI scheme in combination with augmented AO basis sets. For example, an augmented basis set can be obtained from the original AO set by adding some special basis functions. Potentially, the use of such combinations might lead to inconsistent results. This possibility is related to the fact that the standard published auxiliary basis sets have not been developed for fitting the product densities resulting from the augmented basis functions, i.e. they do not contain suitable functions to span the product densities resulting from the added new basis functions.

According to published data one may conclude that the RI-X bases have the same quality as aCD/acCD with a Cholesky decomposition threshold in the range $10^{-4}$ to $10^{-6}$~\cite{aCD1,RIvsCD}. Typically, the Cholesky basis sets are somewhat larger than the corresponding RI-C ones and  , therefore, require a higher computational cost which is a reasonable price to pay for an unbiased and highly accurate auxiliary basis set.

\subsection{Green function method}
Green's function or propagator theory is well established and we just briefly describe it also mentioning some computational details. Most of one-particle propagator methods  are based on the well-known Dyson equation~\cite{OHRN,fetter2003,CEDER77}: 
\begin{equation}
\label{Dyson}
\mathbf{G}(\omega)=\mathbf{G_0}(\omega)+\mathbf{G_0}(\omega)\boldsymbol{\Sigma}(\omega)\mathbf{G}(\omega)\thinspace,
\end{equation}
where $\mathbf{G}(\omega)$ is the one-particle Green's function, $\boldsymbol{\Sigma}(\omega)$ is an energy dependent non-local potential called self-energy and $\mathbf{G_0}(\omega)$ is the \textit{free} Green's function. Formally, the ionization potentials (IPs) and  electron affinities (EAs) of the system under consideration are poles of $\mathbf{G(\omega)}$:
\begin{equation}
\label{DysonSol}
\mathbf{G}(\omega)=\bigl(\boldsymbol{\omega\cdot1}-\boldsymbol{\varepsilon}-\boldsymbol{\Sigma}(\omega)\bigr)^{-1}\thinspace,
\end{equation}
where $\boldsymbol{\varepsilon}$ is the diagonal matrix of the canonical HF one-electron energies and $\boldsymbol1$ is a unit matrix. For convenience, the poles of $\mathbf{G(\omega)}$ can be found from
\begin{equation}
\label{DysonSol2}
\mathrm{det}\bigl(\boldsymbol{\omega\cdot1}-\boldsymbol{\varepsilon}-\boldsymbol{\Sigma}(\omega)\bigr)=0\thinspace.
\end{equation}

The self-energy $\boldsymbol{\Sigma}(\omega)$ itself also possesses a spectral representation~\cite{CEDER77} and this can be used to solve Eq.~(\ref{DysonSol2}) by solving an eigenvalue equation. If the elements of $\boldsymbol{\Sigma}(\omega)$ are explicitly known, the solution of the Dyson equation becomes particularly compact~\cite{Arrow}. In this work we resort for simplicity of representation to the so-called \textit{quasi-particle} approximation, in which the self-energy is diagonal and Eq.~(\ref{DysonSol2}) reduces to:

\begin{equation}
\label{qpapp}
\displaystyle\omega=\varepsilon_q+\Sigma_{qq}(\omega)\thinspace,
\end{equation}
where $q$ is the MO index related to the IP or to the EA we are looking for. In practice Eq.~(\ref{qpapp}) is solved iteratively via the Newton-Raphson method
\begin{equation}
\label{iter}
\omega_q^{(n+1)}=\omega_q^{(n)}+\bigl(\varepsilon_q+\Sigma_{qq}(\omega_q^{(n)})-\omega_q^{(n)}\bigr) P_q^{(n)}\thinspace,
\end{equation}

where $P_q$ is the so-called \textit{pole strength}:
\begin{equation}
 \label{ps}
\displaystyle P_q^{(n)}=\biggl(1-\frac{\partial\Sigma_{qq}(\omega)}{\partial\omega}\biggr\rvert_{\omega=\omega_q^{(n)}}\biggr)^{-1},\thinspace \displaystyle0\le P_q \le1\thinspace.
\end{equation}
The iterative procedure usually starts from a HF one-electron energy of the $q$th MO ($\omega_q^{(0)}=\epsilon_q$) and continues until the absolute difference between previous and current values of a pole is smaller than some given threshold, say $\Delta=10^{-5}$:
\begin{equation}
\label{NR}
\lvert \omega_q^{(n+1)}-\omega_q^{(n)} \rvert\le\Delta\thinspace.
\end{equation}

There are several successful approximations to $\boldsymbol{\Sigma}(\omega)$ in the
literature~\cite{CEDER75,CEDER83,CEDER84,NDADC,NR1,NR2,P23}. Here, for simplicity of presentation we employ the well-known expression of second-order perturbation theory. In the second order, the $q$th diagonal element of the self-energy reads
\begin{equation}
\label{SE}
\Sigma^{\left( 2\right) }_{qq}\left( \omega \right) = \sum\limits_{\substack{i,j\\a}}\frac{\left(\thinspace2({qi\lvert aj})-({qj\lvert ai})\thinspace\right)({qi\lvert aj})}
{\omega+\varepsilon_a-\varepsilon_i-\varepsilon_j}+{\sum\limits_{\substack{i\\a,b}}\frac{\left(\thinspace2({qa\lvert ib})-({qb\lvert ia})\thinspace\right)({qa|ib})}
{\omega+\varepsilon_i-\varepsilon_a-\varepsilon_b}}\thinspace,
\end{equation}
where $a,b$ denote virtual, and $i,j$  occupied spatial HF orbitals. The second order contains the most relevant relaxation and correlation corrections to Koopman's theorem~\cite{Pickup} and self-energy provides convenient checks of new computer codes and the computational experience necessary to implement more general approximations.

\section{Computational details}
All calculations presented here are on C$_{60}$ which is an ideally suited object as it contains many atoms and is of general interest documented by numerous investigations. All calculations of ground  and ionic states of $\mathrm{C_{60}}$ were performed in $\mathrm{D_{2h}}$ symmetry at the experimental gas-phase geometry:\- $R_{CC}=1.458\thinspace\mathrm{\AA{}}$ and $R_{CC}=1.401\thinspace\mathrm{\AA{}}$~\cite{C60GEOM}. The AO basis sets used in this work are Dunning's cc-pVXZ (X=D,T) basis sets ~\cite{CC} and the respective total number of basis function are 840 and 1800. Throughout, the spherical representation of the d- and f-basis functions was used. The calculations with CD were performed by using the MOLCAS quantum chemistry program ~\cite{MOLCAS,ALG3}. The DF/RI MP2 calculations were carried out using the TURBOMOLE~\cite{TURBOMOLE2,DSCF,RIDFT,RIMP2} quantum chemistry package with suitable RI-C and RI-JK auxiliary basis sets corresponding to the original set AOs~\cite{RIC2}. Both CD and DF/RI calculations were done on {Intel\textregistered} {Xeon\textregistered} E5440 (2.83GHz) and AMD {Opteron\texttrademark} 2220 (2.80GHz) based supercomputers~\cite{BWGRID,HELICS}. Fully direct HF and MP2 calculations were carried out with the PC GAMESS/Firefly  program suite ~\cite{PCGAMESS1,PCGAMESS2}. For open shell calculations at the HF level the restricted open-shell approach (ROHF) has been employed~\cite{ROHF}.

For calculating IPs and EAs in the quasiparticle approximation with $\Sigma^{\left(2\right)}_{qq}$ we developed a parallel program called $\mathrm{P{-}RICD}\Sigma$. As input data the $\mathrm{P{-}RICD}\Sigma$ uses integral tables in AO representation and SCF MO LCAO coefficients which are generated with the MOLCAS program. Execution of $\mathrm{P{-}RICD}\Sigma$ consists in two separate steps: parallel transformation of integral tables from the AO to the MO representation by Eq.~(\ref{transformation}) and iterative solving Eq.~(\ref{iter}). The ERIs needed during the iterative solution are recomputed in parallel by formula~(\ref{CD}). More details about the structure and parallelization of $\mathrm{P{-}RICD}\Sigma$ will be the subject of a forthcoming manuscript. 

All electron propagator calculations in the present paper were fully correlated, i.e. all electrons were taken into account at
the $\Sigma^{\left(2\right)}_{qq}$ level of theory. For the energy conversion of units the factor 1 hartree $\mathrm{(E_H)}=$27.211396 eV was used.

\section{Results and discussion}
First, we would like to introduce the abbreviations used in this chapter. CD-n refers to the full-CD decomposition with threshold $\delta=10^{-n}$. aCD-n* and acCD-n* mean "\textit{atomic Cholesky}" basis and its compact form, respectively. aCD-n or acCD-n basis sets have been formed from the original
aCD-n*/acCD-n* ones by  removing the highest angular momentum orbitals. The term "\textit{low Cholesky}" stands for the full-CD results obtained with a decomposition threshold $\delta$ in the range from $10^{-4}$ to $10^{-6}$, and "\textit{medium Cholesky}" and "\textit{high Cholesky}" are for thresholds in the range $10^{-6}$ to $10^{-8}$ and $10^{-8}$ to $10^{-10}$, respectively.
\subsection{Ground state (GS)}
We would like to start the discussion with MP2 ground state energies because the expressions for MP2, Eq.~(\ref{MP2}), and $\Sigma^{\left(2\right)}_{qq}$, Eq.~(\ref{SE}), are quite similar. For the sake of convenience and clarity, we decompose the MP2 total energy into a sum of two contributions
\begin{equation}
 \label{MP2D}
  \displaystyle E_{Total}^{MP2}=E_{SCF}+E_{MP2_C},
\end{equation}
where $E_{SCF}$ and $E_{MP2_C}$ are the HF energy and the MP2 correlation energy, respectively. With the decomposition given in Eq.~(\ref{MP2D}), the total MP2 energy error takes on the form
\begin{equation}
 \displaystyle{\epsilon_{MP2}}=\epsilon_{HF}+\epsilon_{MP2_C},
\end{equation}
where $\epsilon_X=\lvert E_X^{Approx}-E_X^{Direct}\lvert$ for $\mathrm{X=HF,MP2_C}$.

Figure 1 shows the total MP2 energy error ${\epsilon_{MP2}}$ and its contributions $\epsilon_{HF}$ and $\epsilon_{MP2_C}$. First of all, Figure 1 clearly represents the accuracy of approximations used: the high Cholesky results approach the exact one, DF/RI and the low Cholesky results are the least accurate and atomic CD results are between them. Within the atomic CD series, one can see that the aCD*/acCD* results are on the high-accuracy side, whereas its optimized (reduced) aCD/acCD version are on the lower-accuracy side. We should stress that the observed tendencies are in full agreement with previously published data~\cite{aCDCal}. 

The next important issue concerns the convergence rates of the energies of the full-CD approximation. From Figure 1, it becomes evident that the main source of error here comes from the HF method. We now analyze this trend in more detail. Table I reports the total MP2 energies calculated by using full-CD and DF/RI approximations. By comparing results from the 3-rd and 4-th columns or from the 6-th and 7-th columns with the reference numbers, we conclude that the correlation  energy $(MP2_C)$ converges much faster than the corresponding HF energy $(SCF)$. We attribute this behaviour to the fact that the HF total energy is an expectation value, while the MP2 correlation energy is a correction quantity. Using CD or DF/RI leads to loss in accuracy of the computed ERIs and it is well-known that computing corrections is numerically more robust with respect to the precision of input data, than computing expectation values  due to cancellation of errors.

In view of the results of this chapter we expect propagator methods to be robust with respect to Cholesky decomposition thresholds.

\subsection{Cationic and anionic states}
The GS of $\mathrm{C_{60}}$ is $^1\mathrm{A_{1g}}$. Removing a single electron  yields an $^2\mathrm{H_u}$ cationic
ground state, while the attachment of an extra electron yields a  $\mathrm{^2T_{1u}}$ anionic
ground state. The anion $\mathrm{C_{60}^-}$ is known to be bound  by 2.68 eV in the gas phase~\cite{C60EA2}. The first adiabatic ionization energy of $\mathrm{C_{60}}$ has been estimated to be 7.64 eV~\cite{C60IP1,C60IP2,C60IP3,C60IP4}.

Table II lists the results for the first vertical IP and first EA obtained by two uncorrelated approximations: by applying Koopmans Theorem (KT)~\cite{Koopmans} and by the Delta-SCF ($\Delta\mathrm{SCF}$) method~\cite{DELTASCF}. KT results are obtained from the HF calculations on the neutral fullerene. The IP is obtained from the energy of the HOMO and the EA - from the energy of the LUMO. $\Delta\mathrm{SCF}$ refers to the difference  between the total HF energy of neutral and that of its ions. From Table II, we can see that the calculated quantities depend only slightly on the decomposition threshold. The maximal error introduced by the CD approximation is 1 $m$eV or less. In contrast to the HF total energy (see Table I), the $\Delta\mathrm{SCF}$ results converge much faster. For a threshold of $\delta{=}10^{-5}$ or less, the error in the computed IP and EA is negligible. This finding clearly verifies the predicted robustness of relative quantities to the decomposition threshold.

In Table III and Table IV, we present main results of the present paper. These tables report the calculated first IP and EA  employing various Cholesky decompositions and valence basis sets at the $\Sigma^{(2)}_{qq}$ level. As one would expect, the computed quantities are not very sensitive to the Cholesky decomposition threshold. Even the low Cholesky results have a very encouraging level of accuracy of 1 $m$eV. Starting from $\delta{=}10^{-6}$, the medium Cholesky results approximate the exact ones ($\delta{=}10^{-10}$) fairly well. Clearly, the full Cholesky basis sets provide superior convergence in calculating first IPs and EAs.

Interestingly, the atomic CD results resemble the full-CD ones. From Table III and IV, it is clearly seen that here there is no difference in the results obtained using original atomic Cholesky basis or its compact form (aCD-n vs acCD-n or aCD-n* vs acCD-n*). In the case of atomic CD, the most pronounced changes occur by removing higher orbital products, i.e. in going from aCD-n* to aCD-n or from acCD-n* to acCD-n Cholesky basis sets. Within aCD-n/acCD-n Cholesky basis sets, the maximal error in the first IP/EA potential is 2 $m$eV.

In order to  investigate in more detail the influence of Cholesky basis sets on the calculated IP and EA, we extend the energy window of the calculated potentials. Now, all canonical HF orbitals lying in the energy range from -14.272 to 4.088 eV in the cc-pVDZ basis set, and from -14.303 to 9.903 eV in the cc-pVTZ basis set, are taken into account, and only poles of the propagator which are related to quasiparticles are considered in the subsequent error analysis. Before discussing further, we would like to
identify two issues: the reference results and the precision of the statistical results. In the error analysis of the computed spectrums the $\Sigma_{qq}^{\left(2\right)}$/{CD-10} results  are used as reference. Since we use $\Delta=10^{-5}$ as the convergence threshold  in the $\Sigma_{qq}^{\left(2\right)}$ iterative procedure~(\ref{NR}), all computed statistical characteristics which are below the given threshold should be considered zero. The reference spectra (IPs, EAs and their pole strengths as a function of energy) of C$_{60}$ obtained with the cc-pVDZ and cc-pVTZ valence basis sets are shown in Figure 2. While calculations using propagators are available for the IPs even beyond second order self-energies ~\cite{C60Ortiz}, no ab initio calculations on the EAs of C$_{60}$ have been reported so far because of the large computational effort involved. Experimental IPs and EAs are available in literature, see e.g.~\cite{C60IP1,C60IP2,C60IP3,C60IP4,C60EA2}. In Figure 2, we see that enlarging the basis set shifts the EAs and IPs to lower energies by an amount which is approximately constant for each of these groups of quantities. The second order understimates the first IP and inclusion of higher order corrections is essential for a quantitative prediction (see OVGF and ADC(3) results in Table II). In this work, we are not concerned with the absolute quality of the self-energy
used. We are rather concerned with the accuracy of the calculations using a given self-energy and the CD technique. In addition, utilizing the CD enables us to attack larger molecules which cannot computed otherwise.

We now turn to the error analysis of the various CD approximations used. The data shown in Figure 3 and Figure 4 confirm the
observed feature of the full Cholesky basis sets discussed above. As in the case of the first IP and EA, the full-CD spectral results for $\delta=10^{-6}$ or less are exact within the prescribed accuracy limit (10 $\mu\mathrm{E_h}$ or 0.27 $m\mathrm{eV}$). From the figures, it is also evident that the full-CD results depend on the decomposition threshold, but
they are only slightly affected by the quality of the AO basis set used. In other words, full-CD basis sets introduce nearly uniform spectral errors which only slightly depend on the AO basis set used. In particular, the RMAX (RMS) value for CD-4 are 265 (81) and 262 (67) $\mu\mathrm{E_h}$ in the cc-pVDZ and cc-pVTZ basis sets, respectively. Consequently, low Cholesky results have a reasonable accuracy of a few $m$eV, a finding which is of practical relevance.

In contrast to the full-CD results, the atomic CD ones are sensitive to the quality of the AO basis set used. By going from  double- to triple-zeta-quality AO basis sets, we substantially decrease the overall error (RMAX and RMS are reduced by factor of 7) of the spectra computed using \mbox{aCD-n/acCD-n} (n=4,6) Cholesky basis sets. In the case of \mbox{aCD-n*/acCD-n*} Cholesky basis sets the improvements are not so pronounced. In the triple-zeta basis set, the aCD-n*/acCD-n* (n=4,6) results are only 1.07 to 3.63 times better than in the double-zeta set. For cc-pVTZ AO, the \mbox{aCD-4*/acCD-4*} and \mbox{aCD-6/acCD-6} Cholesky basis sets provide sub-milli-electron-volt accuracy, the corresponding value of RMAX  and RMS are about 18 and 5 $\mu\mathrm{E_h}$, respectively. We note that the aCD-6*/acCD-6* results in both AO basis sets used coincide with the reference ones.

Interestingly, the aCD-4/acCD-4 results in the cc-pVTZ AO basis set have milli-electron-volt accuracy, which is much better than the accuracy of the CD-4. For instance, the RMAX (RMS) value in the cc-pVTZ basis set for aCD-4/acCD-4 and CD-4 are about 46 (16) and 262 (67) $\mu\mathrm{E_h}$, respectively. 

In order to characterize the origin of errors in the spectra due to CD, we have made a linear regression analysis between the reference results (CD-10) and the others. The resulting correlation coefficients are 1.00000.  Figure 5 displays a typical picture resulting from the regression analysis. Obviously, the error due to CD is systematic. Therefore, the CD approach leads essentially to a uniform shift of the whole spectrum. Even in the worst case (CD-4) addressed in Figure S1 in the supplementary material, this shift amounts only to 1 $m$eV.

\subsection{Efficiency}
In order to demonstrate the computational power of the developed $\mathrm{P{-}RICD}\Sigma$ program we would like to provide some timings. As is well-known one of the main bottlenecks of quantum chemistry is the transformation of ERIs from AO to MO representation. For the $\mathrm{P{-}RICD}\Sigma$ program running on 150 cores (50 nodes x 3 cores) the typical timings of the AO to MO transformation  in the cc-pVTZ valence basis set range from 20s to 100s within the full-CD series and from 9s to 25s in case for the aCD ones. By using the same number of cores the wall time needed to pass one NR iteration~(\ref{NR}) is between 15s and 25s. Recall that the total number of basis functions in cc-pVTZ for C$_{60}$ is 1800.

Another important point is the timing for computing integral tables via the CD technique. Figure 6 depicts the time needed to complete CD in various Cholesky and valence basis sets. As one can see, the aCD computational scheme is at least one order of magnitude faster than the full-CD ones. In particular, the CD-10 is about 100 times slower than aCD-4.

\section{Conclusions}
In the present work we demonstrate the robustness of the one-particle electron propagator method with respect to Cholesky decomposition schemes for two-electron integrals. All decomposition schemes reported in the literature are used. We found that even for moderate Cholesky decomposition thresholds ($\ge10^{-5}$) the maximal error in computed electron affinities and ionization potentials is rather small ($\sim$1 $m$eV) and is typically several orders of magnitude smaller than the error arising from the incompleteness of the AO basis sets used. The full Cholesky decomposition exhibits excellent convergence properties with respect to the decomposition threshold. For electron propagator methods there is no need to use small ($\le10^{-7}$) thresholds. The atomic Cholesky basis sets speed up the calculations by several orders of magnitude without leading to a significant loss in accuracy. In particular, we conclude that acCD-n and acCD-n* (n=4,6) Cholesky basis sets provide optimal compromise between performance and accuracy.

The error introduced by the Cholesky decomposition has a systematic behavior. Varying the decomposition threshold leads to a nearly uniform shift of the energy of the whole spectrum, i.e. all calculated poles are shifted by about the same value.

We want to stress that the results presented could be obtained in a reasonable time only because of the efficient parallel algorithm employed and by utilizing a massive parallel computer. By using the Cholesky decomposition
technique and parallel computing one is now able to perform large-scale electron propagator calculations, which were impossible before via conventional techniques. This opens up wider perspectives in modelling large molecular systems.

\section{Acknowledgements}
Financial support by the Deutsche Forschungsgemeinschaft (DFG) is gratefully acknowledged. V.V.P. is grateful to F. Aquilante (Geneva University), V. Veryazov and R. Lindh (Lund University) of the MOLCAS team for their help at the stage of development of the interface to Molcas v7. V.V.P. also appreciates A. Streltsov, E. Gromov and A. Dutoi for reading the manuscript prior to publication and their helpful comments.

\begin{thebibliography}{81}
\bibitem[{\citenamefont{SORT}(1996)}]{SORT}
\bibinfo{note}{Without loss of generality we assume that the original product densities of $\mathfrak{R}$ are presorted in descending order: $\left( \rho{_I} \lvert V \lvert \rho{_I} \right) \ge \left( \rho{_J} \lvert V  \lvert \rho{_J} \right)$,$\thinspace\left (I{<}J,\thinspace J\le N\left( N+1\right)/2\right)$; see the notation used in the text, Chapter II.A }.
\end{thebibliography}

\newpage
\begin{center}
{FIGURE CAPTIONS} 
\end{center}

Figure 1: (Color online) Ground state results on C$_{60}$. Total MP2 energy errors and their contributions employing different Cholesky basis sets for cc-pVTZ valence basis set. All errors computed as absolute deviation from the corresponding results of
MP2(FC)/cc-pVTZ fully-direct calculation. Note that the errors are given in $m\mathrm{E_h}$ while the inset is in $\mu\mathrm{E_h}$. The basis set contains 1800 functions. \\

Figure 2: (Color online) The $\Sigma^{(2)}_{qq}$/CD-10 electron spectra of C$_{60}$ computed in the cc-pVDZ and cc-pVTZ valence basis sets. EAs and IPs are shown on the left and right hand sides, respectively.\\

Figure 3: (Color online) Maximal absolute error of the computed $\Sigma^{(2)}_{qq}$ electron spectra of C$_{60}$ employing different Cholesky basis sets for cc-pVDZ and cc-pVTZ valence basis sets. All errors are computed relative to the $\Sigma^{(2)}_{qq}$/CD-10 spectra. The dashed horizontal line displays the predefined level of accuracy (10 $\mu\mathrm {E_{h}}$).\\

Figure 4: (Color online) RMS error of the computed $\Sigma^{(2)}_{qq}$ electron spectra of C$_{60}$ employing different Cholesky basis sets for cc-pVDZ and cc-pVTZ valence basis sets. All errors are computed relative to the $\Sigma^{(2)}_{qq}$/CD-10 spectra. The dashed horizontal line displays the predefined level of accuracy (10 $\mu\mathrm {E_{h}}$).\\

Figure 5: Correlation between $\Sigma^{(2)}_{qq}$/CD-10 and $\Sigma^{(2)}_{qq}$/acCD-4 results for C$_{60}$ in cc-pVTZ valence basis set. In parenthesis is displayed the correlation coefficient (R).\\

Figure 6: (Color online) The relative timings (t) of the Cholesky decomposition performed within different Cholesky and valence basis sets. The aCD-4 timings ($\mathrm{t_{aCD-4}}$) are used as reference.
For 150 cores (50 nodes x 3 cores) those timings are 6.8  and 106 seconds for the cc-pVDZ and cc-pVTZ valence basis sets, respectively.\\

Figure S1: Correlation between $\Sigma^{(2)}_{qq}$/CD-10 and $\Sigma^{(2)}_{qq}$/CD-4 results for C$_{60}$ in cc-pVTZ valence basis set. In parenthesis is displayed the correlation
coefficient (R).

\newpage

\begin{table}[h]
\vspace{1.5ex}
\caption{Total MP2 energies ($E_h$) for the ground state of ${\mathrm{C_{60}}}$ and their decomposition into SCF (HF) and correlation (MP2c) energies. Listed are the results obtained for different DF/RI and full-CD decompositions for two basis sets cc-pVDZ and cc-pVTZ (840 and 1800 basis functions, repsectively). These results are compared with the "exact" ones.}
\begin{center}
\begin{ruledtabular}
\begin{tabular}{ccccc@{\quad}ccc}
\multicolumn{2}{l}{\multirow{2}{*}{Method}}&\multicolumn{3}{c}{cc-pVDZ}&\multicolumn{3}{c}{cc-pVTZ} \\
\cline{3-5}\cline{6-8}
& & SCF & $\rm{MP2_C}$ & MP2 & SCF & $\rm{MP2_C}$ & MP2 \\
\cline{1-2}\cline{3-5}\cline{6-8}
\multicolumn{1}{c}{\multirow{2}{*}{\small{\bf RI}}} & C & -2271.947700 & -7.741105 & -2279.688805 & -2272.396894 & -9.250503 & -2281.647397  \\
& \small{JK-C} & \multicolumn{3}{c}{--} & -2272.395429 & -9.249306 & -2281.644736  \\
\cline{2-5}\cline{6-8}
\multicolumn{1}{c}{\multirow{7}{*}{\bf CD }}
& \small4 & \small-2271.932301 & \small-7.736309 & \small-2279.668610 & \small-2272.390452 & \small-9.249999 & \small-2281.640451 \\
& \small5 & \small-2271.946702 & \small-7.741485 & \small-2279.688187 & \small-2272.396589 & \small-9.251553 & \small-2281.648142 \\
\cline{2-5}\cline{6-8}
& \small6 & -2271.947519 & -7.741778 & -2279.689297 & -2272.396760 & -9.251745 & -2281.648505 \\
& \small7 & -2271.947671 & -7.741839 & -2279.689510 & -2272.396871 & -9.251791 & -2281.648661 \\
\cline{2-5}\cline{6-8}
& \small8  & -2271.947697 & -7.741858 & -2279.689556 & -2272.396891 & -9.251797 & -2281.648689 \\
& \small9  & -2271.947699 & -7.741857 & -2279.689556 & -2272.396894 & -9.251799 & -2281.648693 \\
& \small10 & -2271.947700 & -7.741857 & -2279.689557 & -2272.396894 & -9.251798 & -2281.648692 \\
\cline{2-5}\cline{6-8}
\bf Direct & & -2271.947700 & -7.741856 & -2279.689556 & -2272.396894 & -9.251798 & -2281.648692  \\
\end{tabular}
\end{ruledtabular}
\end{center}
\vspace{1.5ex}
\end{table}
\newpage

\begin{table}[h]
\renewcommand{\arraystretch}{1.15}
\caption{Vertical IP and EA of $\mathrm{C_{60}}$ at the HF level of theory (eV). The last significant digit is underlined here (in comparison with the corresponding results of the direct calculations).}
{\scriptsize
\begin{center}
\begin{tabular}{ccc@{\hspace{12pt}}cc}
\hline\hline
\multicolumn{1}{c}{\multirow{2}{*}{\ \ Method\ \ }} & \multicolumn{2}{c}{IP} & \multicolumn{2}{c}{EA}  \\
\cline{2-3}\cline{4-5}
& KT & $\Delta\mbox{SCF}$ & KT & $\Delta\mbox{SCF}$ \\
\hline
& \multicolumn{4}{c}{cc-pVDZ}\\
CD-4   & 7.80\underline9 & 7.50\underline4 & 0.76\underline7 & 0.99\underline3 \\
CD-5   & 7.810 & 7.505 & 0.768 & 0.994 \\
CD-6   & 7.810 & 7.505 & 0.768 & 0.994 \\
CD-7   & 7.810 & 7.505 & 0.768 & 0.994 \\
CD-8   & 7.810 & 7.505 & 0.768 & 0.994 \\
CD-9   & 7.810 & 7.505 & 0.768 & 0.994 \\
CD-10  & 7.810 & 7.505 & 0.768 & 0.994 \\
Direct & 7.810 & 7.505 & 0.768 & 0.994 \\ 
& \multicolumn{4}{c}{cc-pVTZ}          \\
CD-4   & 7.79\underline7 & 7.45\underline9 & 0.80\underline5 & 1.05\underline9 \\
CD-5   & 7.798 & 7.460 & 0.807 & 1.061 \\
CD-6   & 7.798 & 7.460 & 0.807 & 1.061 \\
CD-7   & 7.798 & 7.460 & 0.807 & 1.061 \\
CD-8   & 7.798 & 7.460 & 0.807 & 1.061 \\
CD-9   & 7.798 & 7.460 & 0.807 & 1.061 \\
CD-10  & 7.798 & 7.460 & 0.807 & 1.061 \\
Direct & 7.798 & 7.460 & 0.807 & 1.061 \\
OVGF & \multicolumn{2}{c}{7.65\footnotemark[1]} &\multicolumn{2}{c}{{\multirow{2}{*}{--}}} \\
ADC(3)\hfill & \multicolumn{2}{c}{7.68\footnotemark[1]} & \\
Exp. & \multicolumn{2}{c}{7.64\footnotemark[2]} & \multicolumn{2}{c}{2.68\footnotemark[3]} \\
\hline\hline
\footnotetext[1]{Reference 88.}
\footnotetext[2]{Reference 82-85.}
\footnotetext[3]{References 81.}
\end{tabular}
\end{center}}
\end{table}
\newpage

\begin{table}[h]
\caption{Vertical IP of $\mathrm{C_{60}}$ at the $\Sigma_{qq}^{\left(2\right)}$ level of theory (eV). The pole strengths are given in parenthesis. The last significant digit is underlined here
(as a reference the CD-10 results are used).}
\renewcommand{\arraystretch}{1.15}
\begin{center}
\begin{tabular}{ccp{0.35in}ccp{0.35in}cccc}
\hline\hline
\multicolumn{1}{c}{\multirow{2}{*}{Method\ }} & \mc{7}{c}{\hspace{2mm}{\small -}$\log\left( {\rm\delta}\right) $} \\ 
\cline{2-8}
 & \small4 & \small{\ \ 5} & \small6 & \small7 & \small{\ \ 8} & \small9 & \small10 \\ 
\hline
\multicolumn{8}{c}{cc-pVDZ (0.802)}\\ 
\vspace{0.1cm}
{\hfill CD\hspace{1.5mm}}    &  6.94\underline{7} & 6.94\underline{7} & 6.948 & 6.948 & 6.948 & 6.948 & 6.948 \\
{\hfill aCD*}   &  6.94\underline{7} & 6.94\underline{7} & 6.94\underline{7} & 6.948 & 6.948 & 6.948 & 6.948\\
\vspace{0.1cm}
{\hfill acCD*}  &  6.94\underline{7} & 6.94\underline{7} & 6.94\underline{7} & 6.948 & 6.948 & 6.948 & 6.948\\
{\hfill aCD\hspace{1.5mm}}                &  6.94\underline{6} & 6.94\underline{7} & 6.948 & 6.948 & 6.948 & 6.948 & 6.948\\
{\hfill acCD\hspace{1.5mm}}               &  6.94\underline{6} & 6.94\underline{7} & 6.948 & 6.948 & 6.948 & 6.948 & 6.948 \\
\multicolumn{8}{c}{cc-pVTZ (0.793)}\\ 
\vspace{0.1cm}
{\hfill CD\hspace{1.5mm}}    &  7.118             & 7.118             & 7.118 & 7.118 & 7.118 & 7.118 & 7.118 \\
{\hfill aCD*}   &  7.11\underline{9} & 7.118 & 7.118 & 7.118 & 7.118  & \multicolumn{2}{c}{\multirow{2}{*}{--\footnotemark[1]}}\\
\vspace{0.1cm}
{\hfill acCD*}  &  7.11\underline{9} & 7.118 & 7.118 & 7.118 & 7.118  \\
{\hfill aCD\hspace{1.5mm}}                &  7.11\underline{9} & 7.11\underline{9} & 7.11\underline{9} & 7.11\underline{9} & 7.11\underline{9} & 7.11\underline{9} & 7.11\underline{9} \\
{\hfill acCD\hspace{1.5mm}}               &  7.11\underline{9} & 7.11\underline{9} & 7.11\underline{9} & 7.11\underline{9} & 7.11\underline{9} & 7.11\underline{9} & 7.11\underline{9} \\
\hline\hline\\
\end{tabular}
\footnotetext[1]{Calculation failed due to some internal restriction of MOLCAS.}
\end{center}
\end{table}
\newpage

\begin{table}[h]
\caption{Vertical EA of $\mathrm{C_{60}}$ at the $\Sigma_{qq}^{\left(2\right)}$ level of theory (eV). The pole strengths are given in parenthesis. The last significant digit is underlined here
(as a reference the CD-10 results are used).}
\renewcommand{\arraystretch}{1.15}
\begin{center}
\begin{tabular}{ccp{0.35in}ccp{0.35in}cccc}
\hline\hline
\multicolumn{1}{c}{\multirow{2}{*}{Method\ }} & \mc{7}{c}{\hspace{2mm}{\small -}$\log\left( {\rm\delta}\right) $} \\ 
\cline{2-8}
 & \small4 & \small{\ \ 5} & \small6 & \small7 & \small{\ \ 8} & \small9 & \small10 \\ 
\hline
\multicolumn{8}{c}{cc-pVDZ (0.819)}\\ 
\vspace{0.1cm}
{\hfill CD\hspace{1.5mm}}    &  2.75\underline{3} & 2.754 & 2.754 & 2.754 & 2.754 & 2.754 & 2.754 \\
{\hfill aCD*}   &  2.75\underline{3} & 2.754 & 2.754 & 2.754 & 2.754 & 2.754 & 2.754 \\
\vspace{0.1cm}
{\hfill acCD*}  &  2.75\underline{3} & 2.754 & 2.754 & 2.754 & 2.754 & 2.754 & 2.754\\
{\hfill aCD\hspace{1.5mm}}                &  2.75\underline{2} & 2.75\underline{3} & 2.75\underline{5} & 2.75\underline{5} & 2.75\underline{5} & 2.754 & 2.754  \\
{\hfill acCD\hspace{1.5mm}}               &  2.75\underline{2} & 2.75\underline{3} & 2.754 & 2.75\underline{5} & 2.75\underline{5} & 2.754  & 2.754 \\
\multicolumn{8}{c}{cc-pVTZ (0.815)}\\ 
\vspace{0.1cm}
{\hfill CD\hspace{1.5mm}}    &  3.11\underline{0} & 3.111 & 3.111 & 3.111 & 3.111 & 3.111 & 3.111 \\
{\hfill aCD*}   &  3.111 & 3.111 & 3.111 & 3.111 & 3.111  & \multicolumn{2}{c}{\multirow{2}{*}{--\footnote{Calculation failed due to some internal restriction of MOLCAS.}}}\\ 
\vspace{0.1cm}
{\hfill acCD*}  &  3.111 & 3.111 & 3.111 & 3.111 & 3.111  \\
{\hfill aCD\hspace{1.5mm}}                &  3.111 & 3.111 & 3.111 & 3.111 & 3.111 & 3.111 & 3.111\\
{\hfill acCD\hspace{1.5mm}}               &  3.111 & 3.111 & 3.111 & 3.111 & 3.111 & 3.111 & 3.111\\
\hline\hline\\
\end{tabular}
\end{center}
\end{table}
\newpage 

\begin{center}

\begin{figure}
\vspace{8.0cm}
\centering
\begin{center}
\includegraphics[angle=0,scale=.90]{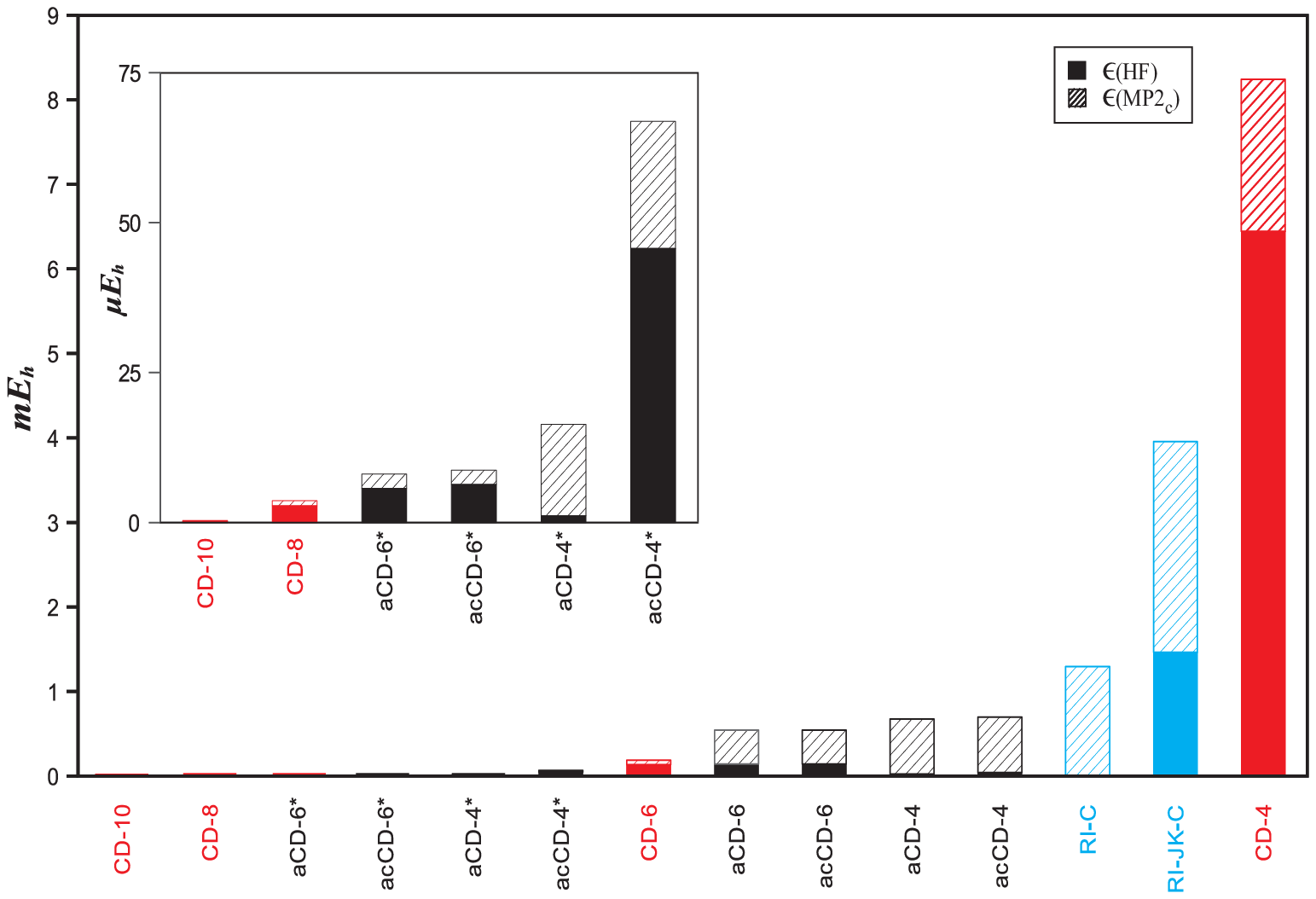}
\end{center}
\caption{}
\end{figure}

\pagebreak

\begin{figure}
\vspace{10cm}
\begin{center}
\includegraphics[angle=0,scale=1.0]{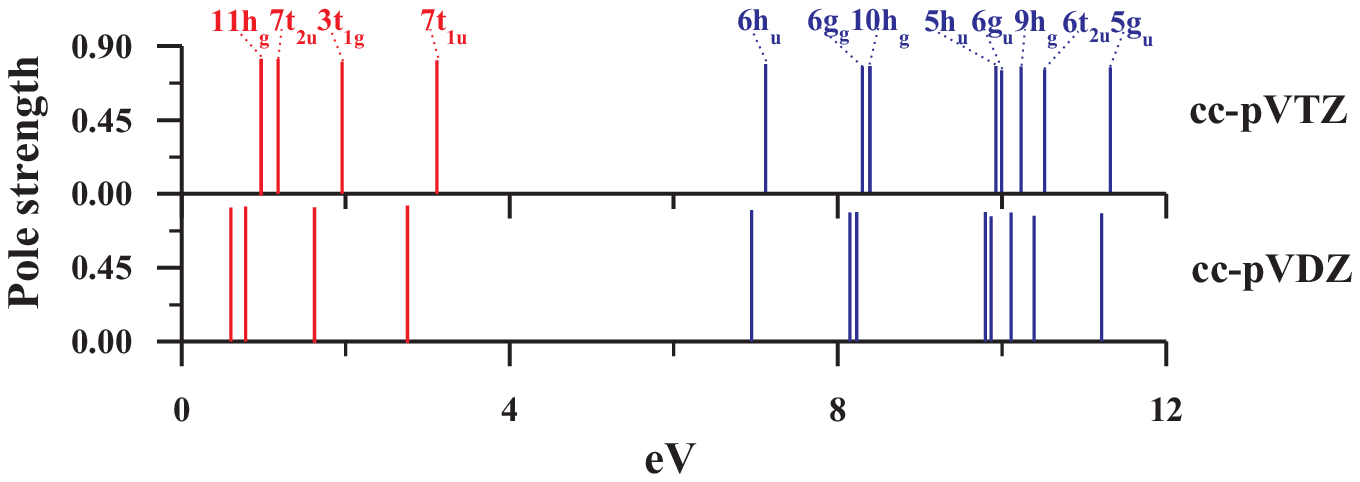}
\end{center}
\caption{}
\end{figure}

\pagebreak

\begin{figure}
\vspace{5cm}
\centering
\begin{center}
\includegraphics[angle=0,scale=1.0]{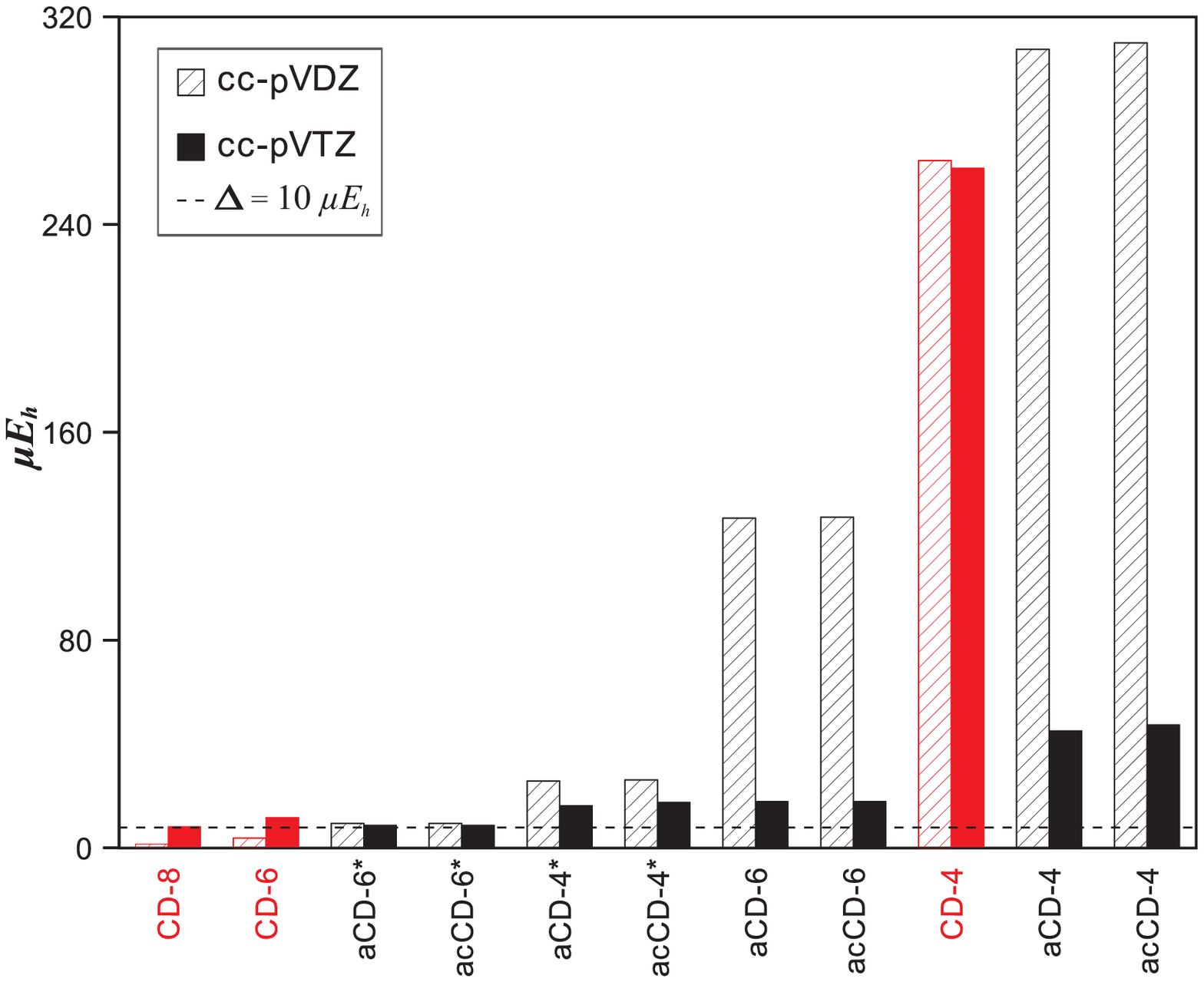}
\end{center}
\caption{}
\end{figure}

\begin{figure}
\vspace{5cm}
\centering
\begin{center}
\includegraphics[angle=0,scale=1.0]{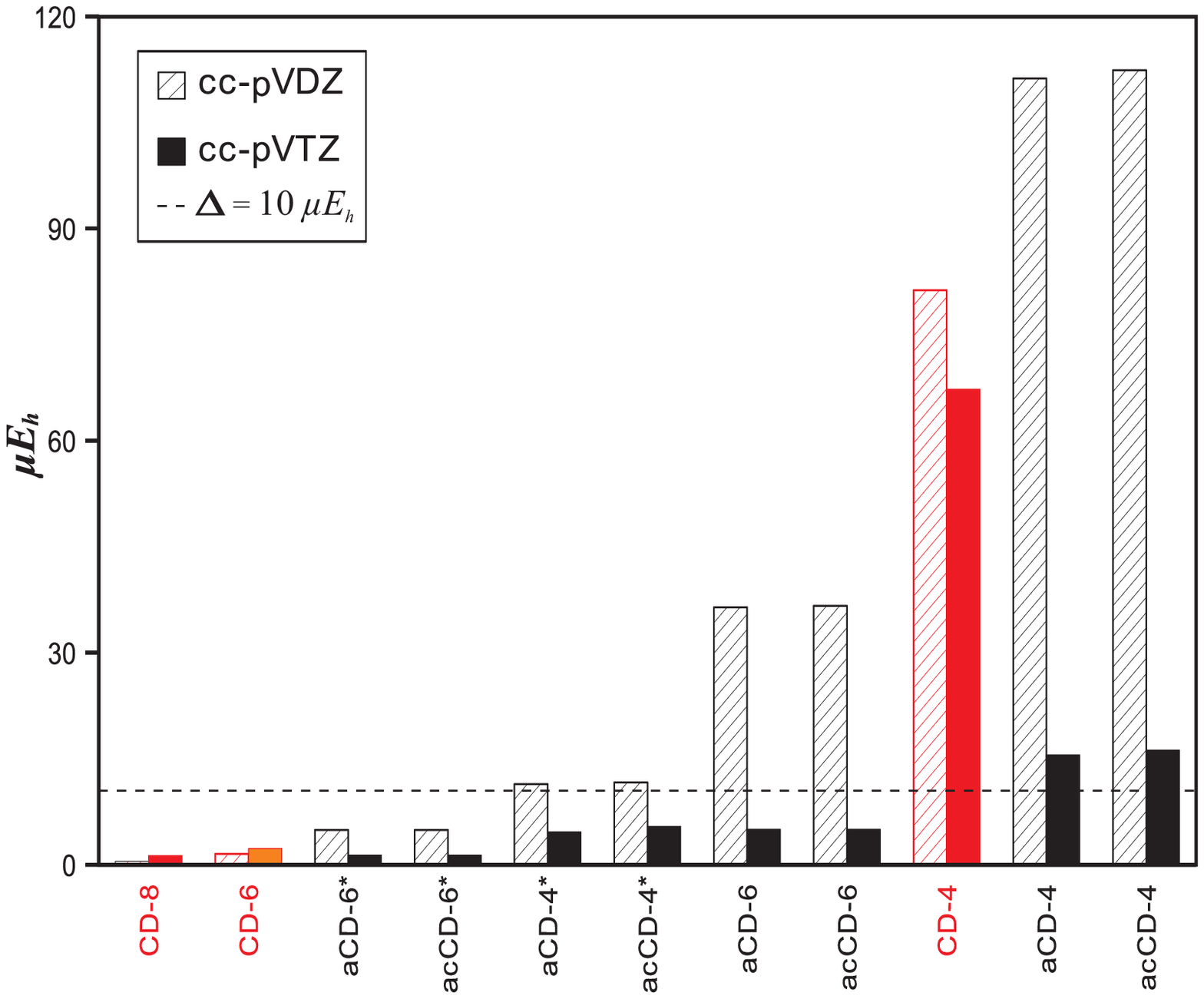}
\end{center}
\caption{}
\end{figure}

\begin{figure}
\vspace{5cm}
\centering
\begin{center}
\includegraphics[angle=0,scale=.80]{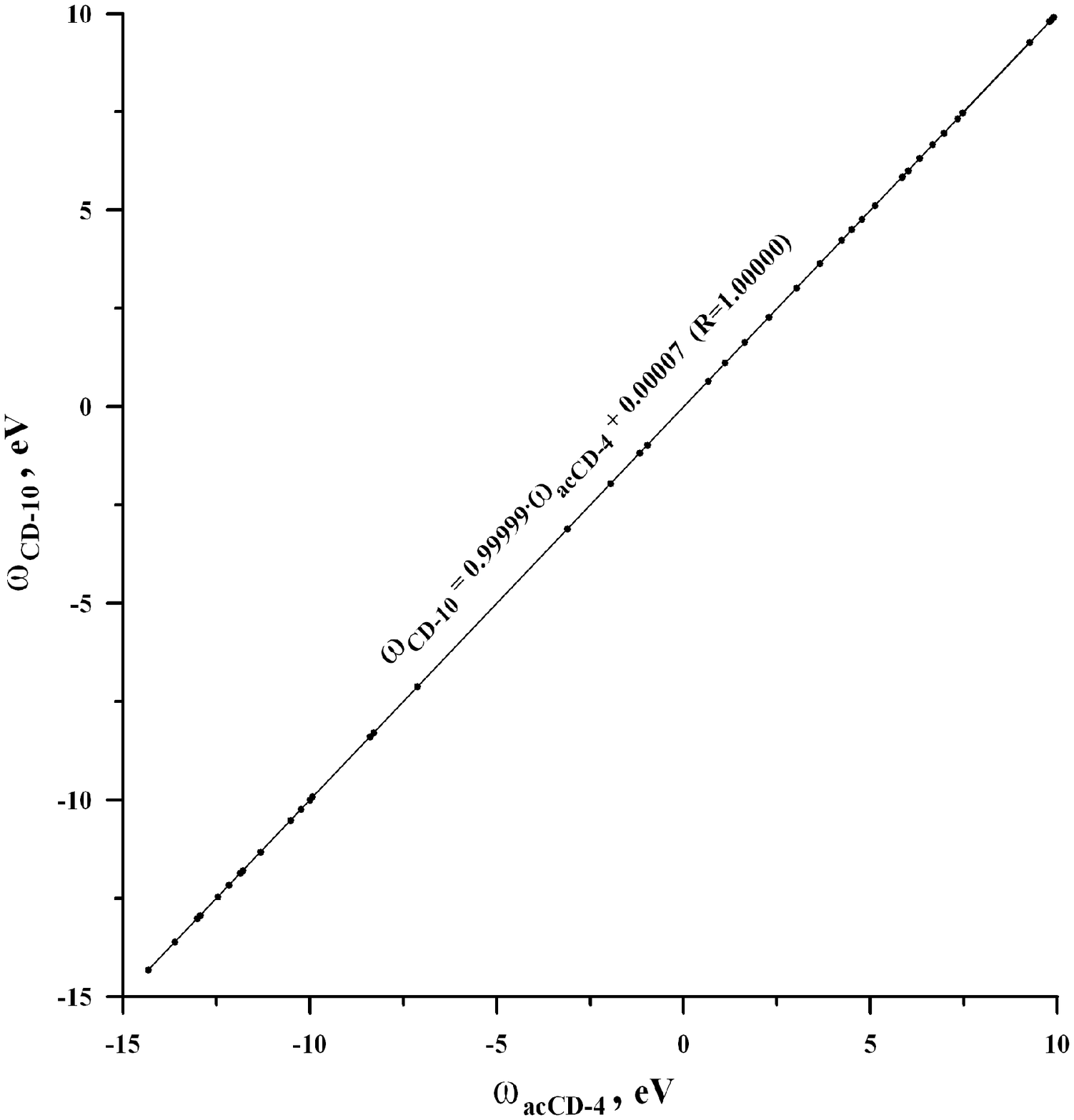}
\end{center}
\caption{}
\end{figure}

\begin{figure}
\vspace{5cm}
\centering
\begin{center}
\includegraphics[angle=0,scale=1.0]{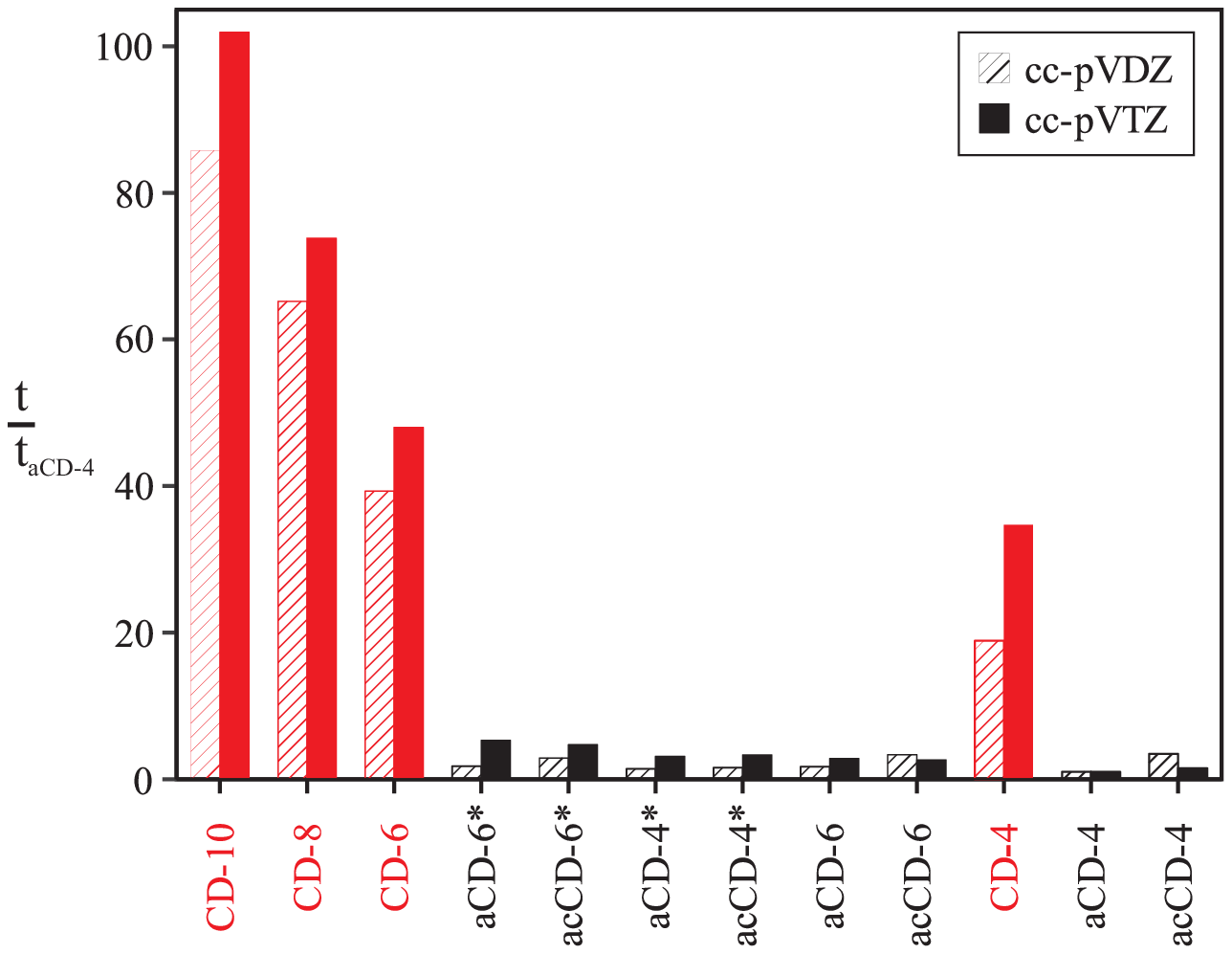}
\end{center}
\caption{}
\end{figure}

\end{center}

\renewcommand{\thefigure}{S1}
\begin{figure}[htp!]
\hspace{-2.5cm}
\begin{minipage}[b]{1.14\linewidth}
\centering
\includegraphics[angle=0,scale=.90]{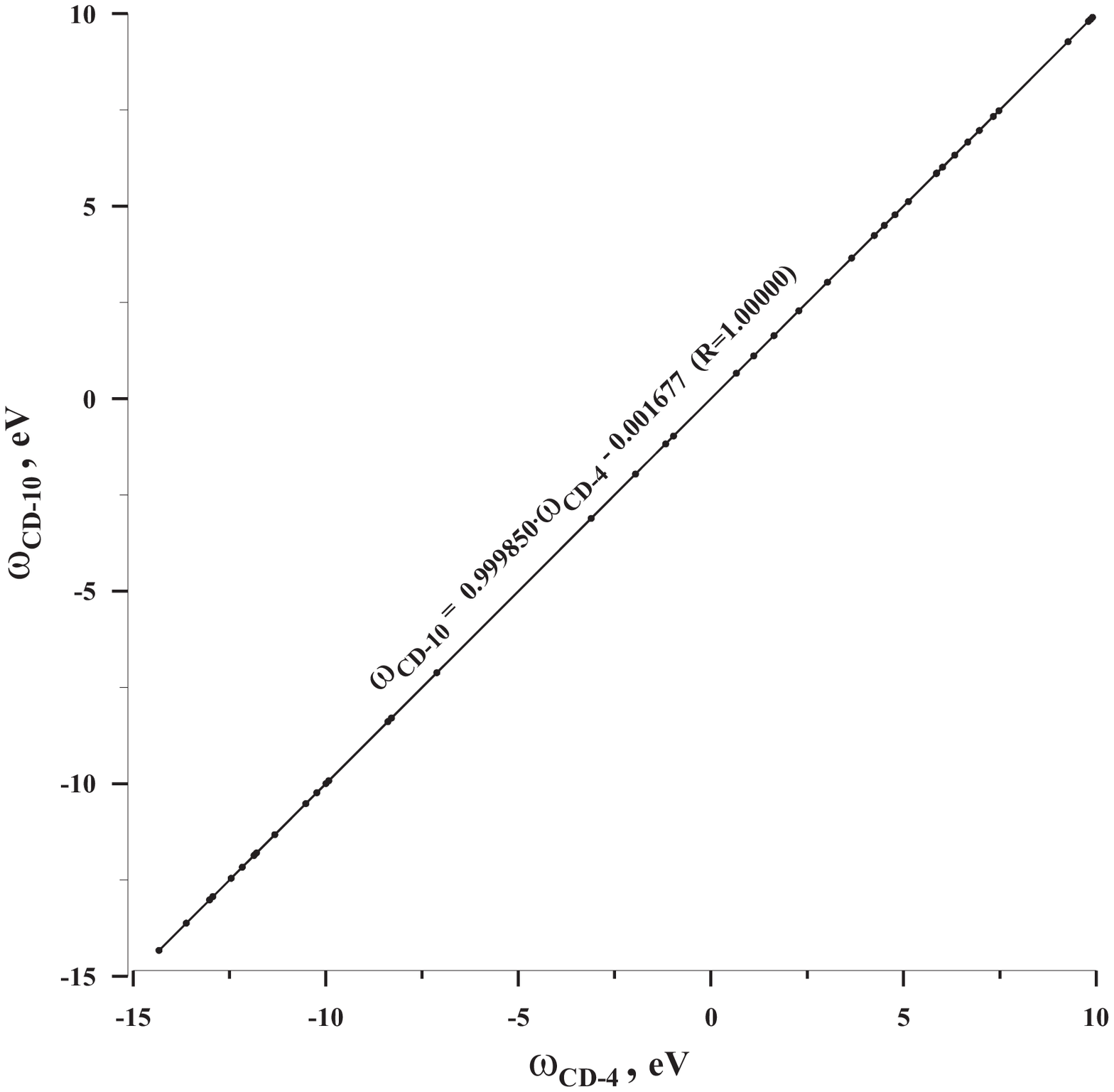}
\caption{}
\end{minipage}
\end{figure}

\end{document}